# Constitutive parameter identification of transtibial residual limb soft tissue using ultrasound indentation and shear wave elastography

Bryan J. Ranger[a,1,2], Kevin M. Moerman[c,1], Brian W. Anthony[d,e], Hugh M. Herr[b]

[a]Department of Engineering, Boston College, 245 Beacon Street, Chestnut Hill, MA 02467 USA
[b]MIT Media Lab, 75 Amherst Street, Massachusetts Institute of Technology, Cambridge, MA 02139 USA
[c]College of Engineering and Informatics, National University of Ireland Galway, Galway, H91HX31, Ireland
[d]Institute for Medical Engineering and Science, 45 Carleton Street, Massachusetts Institute of Technology, Cambridge, MA 02139 USA
[e]Department of Mechanical Engineering, 127 Massachusetts Avenue, Massachusetts Institute of Technology, Cambridge, MA 02139 USA

## Abstract

Finite element analysis (FEA) can be used to evaluate applied interface pressures and internal tissue strains for computational prosthetic socket design. This type of framework requires realistic patient-specific limb geometry and constitutive properties. In recent studies, indentations and inverse FEA with MRI-derived 3D patient geometries were used for constitutive parameter identification. However, long computational times and use of specialized equipment presents challenges for clinical deployment. In this study, we present a novel approach for constitutive parameter identification using a combination of FEA, ultrasound indentation, and shear wave elastography. Local shear modulus measurement using elastography during an ultrasound indentation experiment has particular significance for biomechanical modeling of the residual limb since there are known regional dependencies of soft tissue properties such as varying levels of scarring and atrophy. Beyond prosthesis design, this work has broader implications to the fields of muscle health and monitoring of disease progression.

---

1. Co-first author.
2. Corresponding author. bryan.ranger@bc.edu

## 1. Introduction

For lower extremity amputees, prostheses are commonly used to restore mobility and foster a rapid return to normal physical activities (Tang et al. (2012)). A critical component of the prosthesis is the socket, the cup-like interface that connects to the residual limb (Laing et al. (2011)). Since the socket is the most individual and custom-made component of the prosthesis, its quality plays a major role in determining patient adoption and comfort. For example, poor socket fit can result in residual limb pain which in turn can significantly affect quality of life (Davis (1993)). Further, even mild discomfort is concerning as it may result in limited mobility and an altered gait, which could contribute to long term ailments such as back pain, obesity, musculoskeletal pathologies (e.g., osteoarthritis, osteopenia, and osteoporosis), as well as cardiovascular disease (Gailey et al. (2008); Ma et al. (2014); Tintle et al. (2010)).

Despite how critical the socket is to the prosthesis, it is generally produced using manual plaster casting methods. Alternatively, one means of achieving a quicker and more repeatable production of prosthetic sockets is by using computer aided design/manufacturing (CAD/CAM) systems (Smith and Burgess (2001)). Commercialized CAD/CAM systems provide a means to digitally capture and model the limb, modify the model in a computational environment, and fabricate a socket or socket mold through methods such as 3D printing (Herbert et al. (2005)). Despite its clear disruptive potential, CAD/CAM tools have not reached full clinical efficacy and normally do not inform the design in a data-driven sense since the process remains a manual and experience-based procedure (i.e., a prosthetist must still subjectively modify the model in the computational environment before creating the final socket).

Finite element analysis (FEA) models that capture soft tissue biomechanics allow for detailed evaluation of internal loading conditions such as local soft tissue stresses and strains (Loerakker (2011)). As such, FEA can be utilized as a tool for computational modeling of a residual limb and evaluation of socket designs (refer to Dickinson et al. (2017) for an extensive review of FEA as a tool for modeling the residual limb). One relevant example is that of Portnoy et al. (2007), who demonstrated a real-time and patient-specific FEA method to analyze internal stress in the soft tissues of the residual limb. Their simplified patient-specific FE studies used 2D modeling for simulations of an oblique section of transtibial residual limbs. This method substantially reduced the necessary computational times while maintaining high local spatial resolution. However, macroscopic structural effects are not taken into account and it assumes a linear response of the soft tissue, both of which may not accurately represent the true and often complex constitutive properties of a residual limb and thus its interaction with the prosthetic socket (Papaioannou et al. (2010)). Overall, despite significant progress toward translatable FE models of the residual limb for socket design, such methods are not widely used in clinical settings.

Our lab developed a predictive and patient-specific biomechanical model of the residuum - model geometry and tissue boundaries were derived from magnetic resonance imaging (MRI), while soft tissue non-linear elastic and viscoelastic mechanical behavior was evaluated using inverse FEA of in-vivo indentation experiments (Sengeh et al. (2016)). Despite encouraging results, we outline limitations with our current approach: (i) the high number of iterations involved in inverse FEA, combined with the use of full 3D geometries, has resulted in long computational times, making it difficult for rapid iterations and turnaround between patients; (ii) both MRI and advanced mechanical indentation are costly and not always accessible, particularly in more remote regions or resource-constrained environments; and (iii) regional dependence of soft tissue properties is not taken into account.

To address these challenges, we aim to establish methods for *in vivo* tissue mechanical property characterization in a way that is: (i) readily available for use in a variety of clinics, even those with more constrained resources, (ii) quick (i.e., does not increase consultation time between patient and clinician), (iii) reliable and accurate, (iv) cost-effective so that the price does not increase from standard methods, and (v) straightforward in operation for the user. To this end, this study presents the use of an ultrasound-based indentation system capable of shear wave elastography (SWE) imaging combined with FEA modeling (implemented using GIBBON (Moerman (2018)) and FEBio (Maas et al. (2012))) to identify soft tissue constitutive parameters of a residual limb. Ultrasound was chosen as the imaging modality of choice due to its inherent ability to address the requirements listed above, and has been shown in other studies to be a useful tool to non-invasively identify mechanical properties of the human lower limbs (Frauziols et al. (2016) and Fougeron et al. (2020)).

In addition, many clinical ultrasound scanners are equipped with SWE capabilities, a technique that has shown significant potential for characterizing local stiffness of musculoskeletal tissues (Drakonaki et al. (2012)). Local shear modulus measurement using SWE during an ultrasound indentation experiment has particular significance for biomechanical modeling of the residual limb since there are regional dependencies of soft tissue properties (e.g., varying levels of scarring and atrophy (Henrot et al. (2000))). Finite element analysis can evaluate large strain and non-linear elastic formulations but does not provide a means to derive spatially varying maps of constitutive behavior. By incorporating ultrasound SWE measurements taken during the indentation experiment, we demonstrate that FEA and elastography may be effectively combined to derive spatially varying maps of constitutive behavior.

In this study, we employ an experimental-numerical approach to analyze residual limb tissue constitutive behavior using ultrasound indentation-based FEA and SWE. Experiments consisted of using an ultrasound probe, capable of performing SWE, that is fitted with a custom force sensor to perform a mechanical indentation at select anatomical locations on subjects residual limbs. Numerically, inverse FEA-based parameter fitting was performed by incorporating ultrasound SWE measurements and minimizing the difference between the experimental and simulated force-displacement curves for each to determine constitutive parameters. Our overarching goal was to explore methods in which computational time, accuracy, and potential for clinical implementation can be leveraged for constitutive parameter identification of residual limb soft tissue for use in FEA models and computational socket design, and demonstrate that SWE may inform constitutive modeling of tissue by providing an experimental measure of spatially varying properties.

## 2. Methods

Following a protocol approved by the MIT Committee on the Use of Humans as Experimental Subjects (COUHES), two amputees (one unilateral, one bilateral) were recruited and consented to participate in this study. This allowed for a total *n*=3 residual limbs. In this study, 3D biomechanical models of the residual limb were created from MRI (geometry). Ultrasound indentation combined with elastography and inverse FEA was used to study the tissue mechanical properties. A detailed overview of the modeling schemes that served as the basis of this work are described in Sengeh et al. (2016).

*2.1. Experimental methods*

*2.1.1. MRI*

MRI was used to acquire patient limb geometry. To perform the scan, the patient was situated prone and feet-first inside a 3 Tesla MRI scanner (Siemens Magnetom Tim Trio 3T, Siemens Medical Systems, Erlangen, Germany). All imaging was performed with a RF body coil wrapped around their residuum; no contact was made between the coil and limb so as to prevent tissue deformation. A T1 sequence was used (TR=2530, TE=3.93, acquisition matrix 176x256, 176 slices, voxel size 1.00x1.00x1.00 mm) for image data acquisition. Since the indentation experiment was conducted outside the MRI environment, to specify the desired indentation sites during imaging, MRI-compatible Beekley PinPoint markers (Beekley Corporation, Bristol, CT) were attached to the skin surface prior to imaging.

*2.1.2. Indentation setup*

A schematic depicting the overall setup with patient orientation is shown in **Figure 1**. The ultrasound indentation system has been described in detail in previous studies (Huang (2017)), and builds upon other systems in the literature that have used ultrasound indetnation to assess biomechanical properties (Zheng and Mak (1996)). The current ultrasound system, pictured in **Figure 2**, differs from the published design only in that it was modified to accommodate a GE Logiq E9 ultrasound system (General Electric, Niskayuna, NY). In summary, the design consists of a load cell (Futek, Irvine, CA) which measures forces applied between the ultrasound probe and imaged body segment. A three-axis accelerometer (Analog Devices, Norwood, MA) is mounted within the shell to measure orientation of the device with respect to gravity. Two DAQ boards (National Instruments, Austin, TX) housed in a shielded electronics enclosure, read the load cell and accelerometer voltages. A LabVIEW virtual instrument (VI) running on a tablet records the force and accelerometer data at a rate of 60 Hz, synchronously with ultrasound image capture. The device was designed to fit the GE Logiq E9 9L linear probe; in order to create a custom fit, the probe shape was acquired using a desktop 3D scanner (NextEngine, Santa Monica, CA). SolidWorks was used to design the clamps and shells (average thickness 2.5 mm) based on the 3D scans, which were printed with ABS plastic using a 3D printer (Stratasys, Billerica, MA).

To collect ultrasound indentation data, the 3D-printed shell that houses the probe and force sensor was placed in a linear guide rig (**Figure 2**). This aligns the ultrasound probe perpendicular to the patient's residual limb at the select anatomical location (i.e., the same location at which an MRI marker was placed). A layer of acoustic coupling gel was placed between the skin surface and the probe during imaging. The individual performing the scan then completed an indentation experiment. Indentation force was incremented in steps of 2N starting at 0N up to a maximum of 26N (or lower if the subject reported discomfort). The applied force was set manually by gently rotating the knob of the linear guide mechanism. Forces were adjusted until the force remained constant at the desired level ±0.1N for 30 seconds, which allowed for the tissue to reach a quasi-static state and for the exclusion of viscoelastic effects during the measurement. Once such an equilibrium was obtained for a given force level, force and ultrasound data were recorded.

*2.1.3. Ultrasound imaging*

Ultrasound image capture was collected simultaneously with the force data for each force increment. Two types of ultrasound data were collected 1) clinical B-mode imaging (Frequency: 9 MHz, Depth: 6 cm), and 2) shear wave ultrasound elastography (SWE).

The SWE technique enables the estimation of local mechanical properties following the application of a known mechanical perturbation (Sigrist et al. (2017)), in this case acoustic pulse induced shear waves. SWE data was collected using a built-in function of the GE ultrasound system (based on the methods outlined in (Song et al., 2012, 2013)). The selected SWE region of interest spanned the width of the ultrasound image, was 2.5 cm deep, and placed at the superficial tissue layers of the limb.

The raw image data describes the locally measured shear wave velocity $c_S$. The latter can be related to the tissue density $\rho$ and shear modulus $\mu$ using:

$$c_S = \sqrt{\frac{\mu}{\rho}}$$
$$\mu = \rho c_S^2 \tag{1}$$

The tissue density is generally assumed to be equivalent to that of water, i.e. 1000 $g/m^3$), and therefore SWE measurements offer a direct means of estimating the local shear modulus on a pixel-by-pixel basis. Since the SWE data was acquired simultaneously with force measurements it enables the study of non-linear elasticity.

For each load level the mean and standard deviation of the measurements were calculated. The mean value for each respective patient were subsequently plotted at each level of transducer force during indentation using a box and whisker plot to portray data distribution around a median value.

*2.1.4. Indentation depth analysis*

The indenter depth for each force increment was derived from the B-mode ultrasound data. For each image set, the distance between the skin boundary and the tibia bone contour were determined. The difference between the initial distance and the distance for each increment thus provided the assessment of indentation depth and indenter displacement. The distances were determined manually using a built-in distance measurement feature of the clinical ultrasound system.

*2.2. Computational Modeling*

*2.2.1. Generation of 3D model geometries and meshes*

The steps used to obtain the 3D model geometries consisted of: 1) segmenting tissue contours from the MRI data, 2) converting the tissue contours to surface meshes, 4) creating solid meshes for the surface meshes, 5) determine indentation location and orientation from markers segmented from MRI, 6) reorienting geometries, 7) cropping the full model around the indentation site.

The methods for creating a full 3D multi-material computational FEA model of a residual limb has been described (Sengeh et al. (2016)). To minimize computational time, our current study follows a similar procedure but incorporates a more localized but still anatomically accurate model as an alternative (**Figure 3** and **Figure 4**). To achieve this, surface contours of the skin and bones were segmented from the MRI data using the GIBBON Matlab toolbox (www.gibboncode.org; refer to Moerman (2018), refer to the *imx* function), and then converted to triangulated surface models. From this data, soft tissue (predominantly skeletal muscle) was modeled as one complex, while bones were represented as rigidly supported voids. Each solid material region was meshed with 4-node tri-linear tetrahedral elements using TetGen (www.tetgen.org; refer to Si (2015), which is integrated within the GIBBON Matlab toolbox. Tri-linear tetrahedral elements, such as employed here, can exhibit volume locking when used for nearly-incompressible materials (De Micheli and Mocellin (2009)). In this study, it is likely that element locking could be avoided given the relatively mild nature of the indentations performed on the research subjects (i.e. relatively low force and non-painful), combined with the fact that the mesh was sufficiently refined. A local model was created so as to only include the portion of the limb that was 40 mm radially extended beyond the size of the ultrasound probe head in each direction.

*2.2.2. Boundary condition specification*

The boundary conditions of the probe indentation simulation for each site were derived directly from the ultrasound experimental displacement data. Therefore, the tissue displacement from indentation for each site in the simulation corresponded directly to the experiment. In addition, a zero-friction sliding interface was assumed between the probe and tissue surface. At each ultrasound indentation site, the central point of the head of the ultrasound probe was placed at the marker location derived from the MRI data. The ultrasound probe geometry, created using computer aided design software given measurements of the probe surface, was meshed using triangular shell elements and modeled as a rigid body. Loading orientation of the ultrasound probe was orthogonal to the tissue surface, which was determined from the mean of the local skin surface normal directions.

*2.2.3. Constitutive modeling*

The constitutive formulation used here is a first order form of the coupled hyperelastic strain energy density function presented in Moerman et al. (2016) (obtained by setting *N* = 1 and *q* = 0.5), which in terms of principal stretches $\lambda_i$ is written as:

$$\Psi = \frac{c}{2m^2}\left(\sum_{i=1}^{3}(\lambda_i^m + \lambda_i^{-m} - 2)\right) + \frac{\kappa'}{2}(J-1)^2 \tag{2}$$

Here *c* and $\kappa^0$ are material parameters with units of stress and denote a shear-modulus like and bulk modulus like parameter respectively, the parameters *m* control the degree of non-linearity. Finally, $J = \lambda_1\lambda_2\lambda_3$ is the volume ratio. In this work the bulk-modulus like parameter $\kappa^0$ is linked to the parameter *c* such that:

$$\kappa' = f_\kappa c \tag{3}$$

Where $f_\kappa$ represents the so-called bulk-modulus factor.

Since equation 2 is a second order Ogden formulation (if $c = c_1/2 = c_2/2$ and $m = m_1 = -m_2$, as described in Moerman et al. (2016)) it is referred to here as an *Ogden type*. If the constraint *m* = 2 is imposed this formulation reduces to:

$$\Psi = \frac{c}{8}\left(\sum_{i=1}^{3}(\lambda_i^2 + \lambda_i^{-2} - 2)\right) + \frac{\kappa'}{2}(J-1)^2 \tag{4}$$

This simpler alternative form is referred to here as a *Mooney-Rivlin type* (since it reduces to a Mooney-Rivlin form if *J* = 1).

Both the Mooney-Rivlin type (equation 4) and the more general Ogden type (equation 2) are explored here. The $c$ parameter for these formulations can be related to the initial shear modulus (equivalent to the Hookean shear modulus) $\mu$ (see Lin et al. (2009)) as:

$$c = 2\hat{\mu} \qquad (5)$$

If $\hat{\mu}$ is the mean shear modulus as assessed from ultrasound elastography for the initial state, prior to indentation, then it is clear how the parameter $c$ can be directly informed from the experimental data. The remaining unknown parameters, to be identified from inverse FEA, are $f_\kappa$ for the Mooney-Rivlin type, and $f_\kappa$ and $m$ for the Ogden type.

*2.3. Inverse FEA for constitutive parameter investigation*

An iterative parameter optimization scheme was developed using the GIBBON Matlab toolbox. Both the Ogden type and the Mooney-Rivlin type formulation features the parameter $c$, as mentioned in the previous section this parameter can be directly obtained from the experimental data and was therefore held fixed during all inverse FEA. The remaining parameters were identified using inverse FEA optimization. The following objective function vector $\mathbf{O}(\mathbf{p})$ was defined:

$$\mathbf{O}(\mathbf{p}) = \mathbf{F}_{exp} - \mathbf{F}_{sim} \qquad (6)$$

Which represents the force difference between the experiment and the simulation since $\mathbf{F}_{exp}$ is the experimental force data vector and $\mathbf{F}_{sim}$ is the simulated force data vector sampled (i.e. interpolated using a piecewise cubic Hermite function) at the same displacement increments. The vector $\mathbf{p}$ is the material parameter vector. For the Mooney-Rivlin type formulation, since $m$ is held constant, the parameter vector is simply:

$$\mathbf{p} = [f_\kappa] \quad (7)$$

In the case of the Ogden type formulation $m$ is also optimized leading to:

$$\mathbf{p} = [f_\kappa \quad m] \quad (8)$$

The optimization scheme employs the Levenberg-Marquardt algorithm (Levenberg (1944), as implemented in the Matlab *lsqnonlin* function).

Optimization was done in a two-step process. The Mooney-Rivlin type optimization was performed first, whereby the initial value $f_\kappa = 100$ was used. Once the Mooney-Rivlin type optimization was completed, the Ogden type optimization was started with an initial value for $f_\kappa$ equal to the optimal value from the Mooney-Rivlin type optimization, and $m = 2$.

For optimization the parameters were normalized and the parameters were constrained to $0.01 < f_\kappa < 1000$ and $1 < m < 100$ The objective function tolerance and the parameter tolerance were both set at 0.001.

## 3. Results

### 3.1. Ultrasound elastography measurements

**Figure 5** shows an example sequence of ultrasound shear wave image data collected at a single indentation point at 2 N incremental forces. The unloaded state at 0 N, depicts the initial condition in which the probe is just touching the limb surface - as evidenced by the visible curvature of the skin, it is noted that there was minimal tissue deformation due to probe contact.

As the force increases incrementally by 2 N up to 22 N, the shear wave velocity measurements increase, as expected.

Plots of mean shear wave speed as a function of force for the four indentation points are displayed in **Figure 6** and **Figure 7**. As seen by the linear curve fit lines, each indentation showed an approximately linear increase in shear wave speed over the indentation. Standard deviation also increased in most cases as a function of force. $R^2$ values for each of the four trend lines is greater than or equal to 0.82, while the second research subject's is greater than or equal to 0.90. In addition, standard deviation increased as a function of applied transducer force.

*3.2. Inverse FEA*

The computational time for 1 simulation of 1 model was about 1 minute 30 seconds on a laptop computer (Dell Precision M6800, i7-4910MQ CPU, 32 Gb RAM). The Mooney-Rivlin and Ogden type optimization required 12 and 24 iterations for optimization respectively.

Inverse FEA, showing the FE models of the indentation sites with tissue deformation during indentation are shown in **Figure 8** and **Figure 9**.

**Figure 10** and **Figure 11** provide plots of the raw force-displacement data collected during the indentation experiment, along with the optimized curve fit that resulted from inverse FEA. As can be seen, in each of the four plots, when using an experimental value for *c* and optimizing for *m* and *κ*, we achieved constitutive parameters (**Table 1, 2 and 3**) that accurately reflect the mechanical behavior of the tissue.

## 4. Discussion

The overarching goal of this study was to explore methods in which computational time, accuracy, and potential for clinical implementation can be leveraged for constitutive parameter identification of residual limb soft tissue for use in FEA models and computational socket design. To address some of these challenges, we employed an experimental-numerical approach to analyze residual limb tissue constitutive behavior using a hybrid ultrasound indentation method with FEA and SWE. In doing so, we demonstrate that SWE maps collected from a clinical ultrasound system can be directly incorporated into FE simulation so that patient-specific material parameters are considered.

Many of the current limitations outlined for our lab's current approach to prosthetic socket design are addressed with this technical development. Ultrasound imaging, a cost-effective modality relative to other imaging techniques, combined with indentation was used as an alternative to more advanced mechanical indentation. In the future, ultrasound may be used for both anatomical imaging as an alternative to MRI (Ranger et al. (2015, 2016, 2017, 2019)), theoretically making our approach available in more remote regions or resource-constrained environments as well. In addition, by integrating a local model for FEA simulation, computational time has decreased.

A novel component of our study is the incorporation of SWE as part of the indentation procedure. In the literature, shear moduli for resting leg muscle has been reported to be on average 9.3 +/- 2.0 kPa (Siracusa et al. (2019)). The shear moduli of the indentation points presented in this study ranged from 7.4 - 11.7 kPa, which falls within reasonable limits compared to the expected values. Our analysis with SWE allows for real-time measurements of the regional dependence of soft tissue properties. As shown in **Figure 6** and **Figure 7**, when ultrasound shear wave image data is collected at an indentation point on a transtibial residuum near the gastrocnemius muscle, shear wave velocity measurements increase as a function of applied transducer force. Since the transducer force compresses the tissue, causing internal tissue strains, observing this relationship in the SWE image is expected (Kot et al. (2012)). However, to the authors knowledge, quantifying this correlation and relating it to tissue biomechanics has not yet been adequately explored. Because shear wave elastography provides an absolute measurement of velocity and is dependent on transducer force, it becomes imperative for diagnosis that this relationship is understood. In a forthcoming publication from our groups, we present data on a larger cohort of patients (including non-amputee subjects) to further quantify and analyze how shear wave velocity measurements and applied transducer forces are related.

The results from inverse FEA, presented in **Figure 10**, **Figure 11** and **Table 1, 2 and 3**, show that adequate constitutive parameters may be achieved for the elastic behavior of the muscle when using a patient-specific experimentally measured value for *c* and optimizing for *m* and *κ*. The parameters are comparable to those found in the literature; in particular, in a study by Sengeh et al. (2016), *c* was determined to be 5.2 *kPa*, *m* ranged from 4.74-4.78, and the bulk moduli were set to $\kappa = 100 \cdot c$ (*kPa*). Table 1 consists of the resulting parameters following the iFEA optimization of indentation point 2 on subject 1. As described, the indentation experiment was designed to collect data of the transtibial residuum near the gastrocnemius muscle. Indentation point 1 for subject 1 was located more proximal to the body near the tibia head, whereas indentation point 2 was in a comparable anatomical location as those performed on subject 2 (left and right), thus offering a more direct comparison for our residual limb models to previous work in the literature and between subjects. When comparing the two subjects, the inverse FEA results show differences in the mechanical properties, which is expected. Differences in tissue properties can be explained, in part, by the nature of residual limb tissue, which when reconstructed post-amputation are not conditioned to tolerate dynamic loading experienced by the socket during regular activities. As a result, it is common that after long term prosthesis wear, muscle may atrophy and include increased intramuscular adipose content (Bramley et al. (2021)).

There are some limitations associated with this proof-of-concept study. For simplicity, the limb was modeled only as two components: (1) soft tissue complex and (2) bones – no skin/adipose layers or other anatomical structures such as tendons were accounted for. In future studies, such structures could be incorporated into the modeling scheme and may improve accuracy. Furthermore, the ultrasound probe was oriented transverse to the muscle fiber direction; effects along the fiber direction were not taken into account. Future studies may benefit from imaging in both the longitudinal and transverse orientations relative to muscle fiber direction. In addition, shear wave elastography is a dynamic measurement technique, but given system limitations of the GE ultrasound machine utilized in this study, we had to make quasi-static assumptions when developing our model. The primary system limitation was the frame rate – to acquire one shear wave velocity map, the frame rate was approximately 1 frame per second. Given this, scan time during an indentation experiment was limited since multiple measurements at each incremental force were needed to ensure accurate and precise measurements were achieved. In future studies, in which a shear wave elastography system has a quicker frame rate, it is possible to use this measurement technique to perform dynamic loading experiments in 2D or 3D.

We also note several sources of potential error in this study. First, the MRI scan performed on the research subjects was not optimized for extremity imaging, thus there was distortion at some anatomical boundaries which could affect the accuracy of the segmented limb geometry. Force displacement curves were acquired manually using built-in measuring functions of the clinical ultrasound system; in the future this process should be automated with intensity-based filtering and tracking, or by tracking the probe position. In addition, for each experiment, we used the first indentation, but no pre-conditioning was taken into account. In future iterations, we suggest moving the transducer into position at each force, waiting 20-30 seconds to allow the tissue equilibrate, and then collect shear wave velocity measurements. Given the duration of the indentation, it is possible that the patient may have shifted; though there were no noticeable artifacts that resulted from patient movement, motion was not compensated for in any way. Despite these potential sources for error, they do not detract from the proof-of-concept demonstrated as part of this study.

There are numerous future studies that could expand the work of the approach presented here, particularly as it relates to biomechanical tissue modeling. For example, ultrasound indentation techniques can be expanded to include viscoelasticity and anisotropy, which would generate an even more robust model of limb muscle for this application. These future studies could build off of previous work in MRI elastography that have incorporated nonlinear anisotropic behavior (Wang et al. (2019); Palnitkar et al. (2019); Guertler et al. (2020); Hou et al. (2020, 2022) into similar computational models. For example, local fiber directions as measured by ultrasound may be included in the finite element model to yield initial constants for anisotropic Hookean forms (e.g., transverse isotropy) which can inform initial moduli of anisotropic large strain formulations. Ultrasound may also be employed to determine the unloaded state of tissue where it is unknown. Many tissues in the human body are naturally in a pre-loaded state, so unknown spatially varying pre-stresses and pre-strains exist. Therefore, if a material region is subjected to known tensile and compressive strains (with respect to some chosen reference frame) while simultaneously multiple elastography data sets are recorded, it is possible to determine a minimum stiffness state. Once a configuration is identified for which a material region presents with the minimum stiffness this configuration may serve, for that region, as the new initial or reference configuration. This configuration could be used to retrospectively adjust the recorded deformations to represent deformations with respect to this configuration as the initial state.

The results of this study could also extend beyond the realm of prosthetic socket design to various clinical applications. For example, it may inform detection, diagnosis and modeling of soft tissue abnormalities, which are often correlated with a local change in tissue stiffness. This has particular relevance for cancerous lesions, which are well-documented to appear as hard nodules relative to their surrounding tissue environment and could be the subject of a future research study.

## 5. Conclusion

In conclusion, we present a hybrid approach to determine soft tissue constitutive parameters that utilizes local shear modulus measurements from elastography during an ultrasound indentation experiment in combination with inverse finite element analysis. This work has particular significance for biomechanical modeling of residual limbs since there are known regional dependencies of soft tissue properties such as varying levels of scarring and atrophy. Beyond prosthesis design, this work has broader implications to the fields of muscle health and monitoring of disease progression.

## 6. Acknowledgements

The authors would like to thank the students and staff of the Computational Instrumentation and Device Realization Lab for their support related to this project; in particular, we would like to extend our gratitude to Rebecca Zubajlo and Athena Huang for their help with ultrasound hardware development and data collection, and Micha Feigin, Xiang Zhang and Jonathan Fincke for their guidance on various aspects of the project. Further, the authors would like to thank Dana Solav, Aaron Jaeger, Arthur Petron, David Sengeh, and the rest of the students and staff of the Biomechatronics Group at the MIT Media Lab for their support and guidance. This work was funded in part by the National Science Foundation Graduate Research Fellowship Program (NSF-GRFP), the MIT Media Lab Consortia, and the Robert Woods Johnson Foundation.

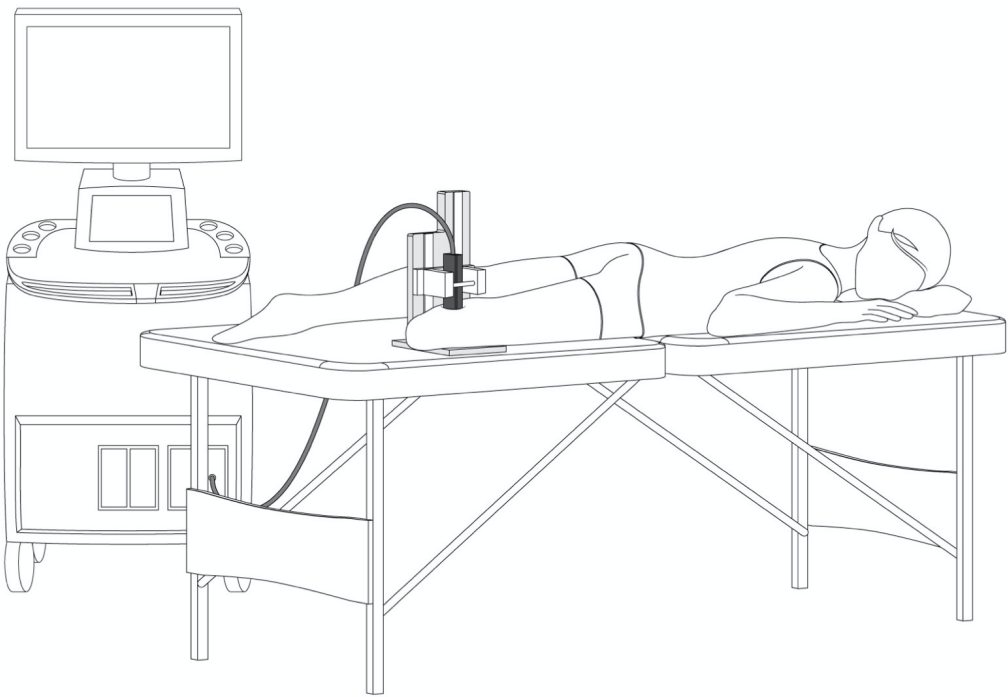

Figure 1: *Schematic showing overall indentation setup and patient positioning for experiments.*

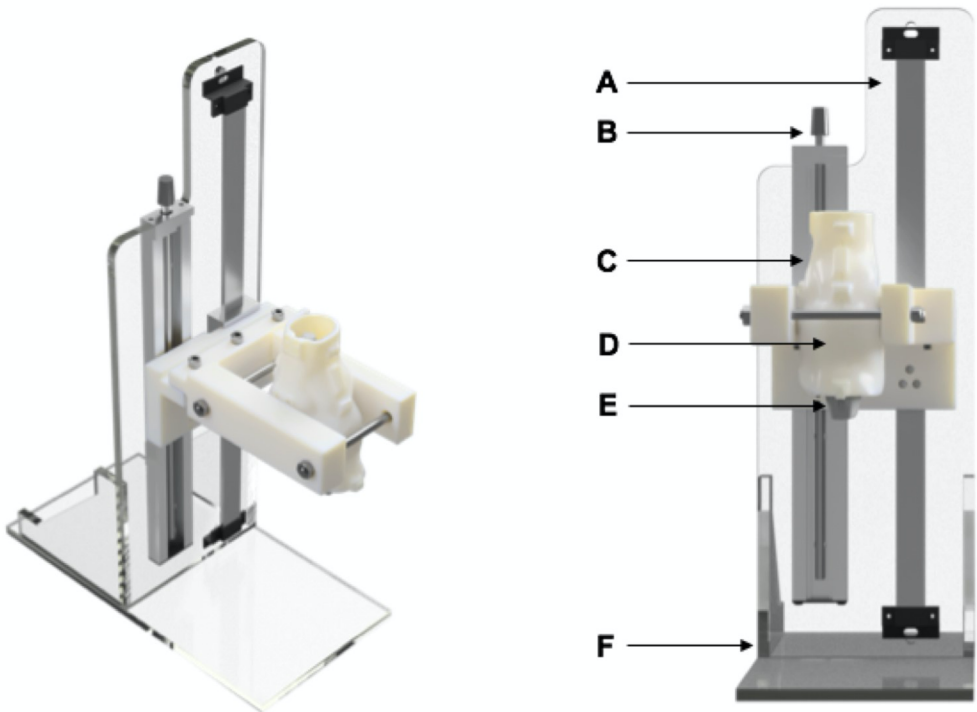

Figure 2: *CAD rendering of ultrasound indentation setup. A Linear guide. B Knob to manually drive movement of the probe vertically for an indentation experiment. C Custom shell that mounts the ultrasound probe to the rest of the system. D Force sensor (located inside of shell, and mounted to the ultrasound probe). E Ultrasound probe. F Acrylic frame.*

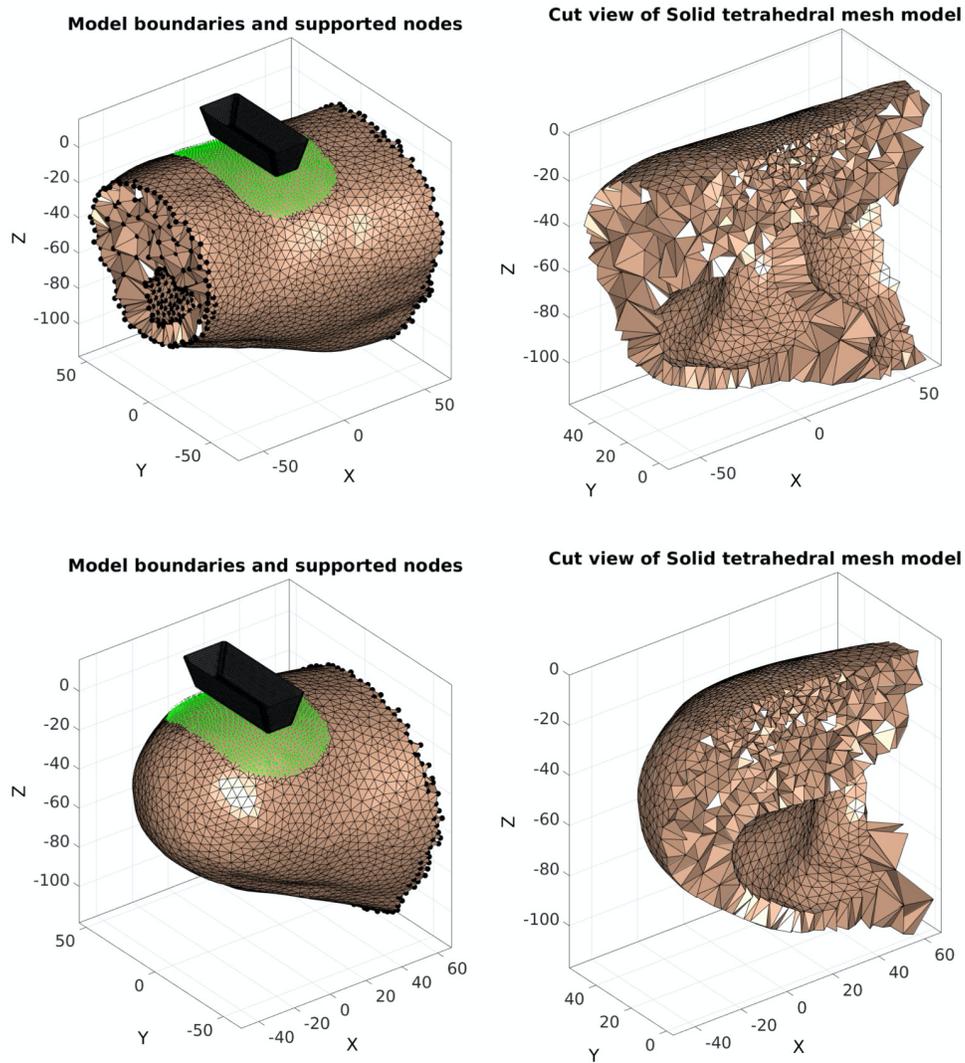

Figure 3: *FE meshes of the anatomically accurate local computational model created from surfaces of bones and skin segmented from the MRI data. This allows for simulation of the ultrasound indentation. (Top) Subject 1, indentation point 1; (Bottom) Subject 1, indentation point 2.*

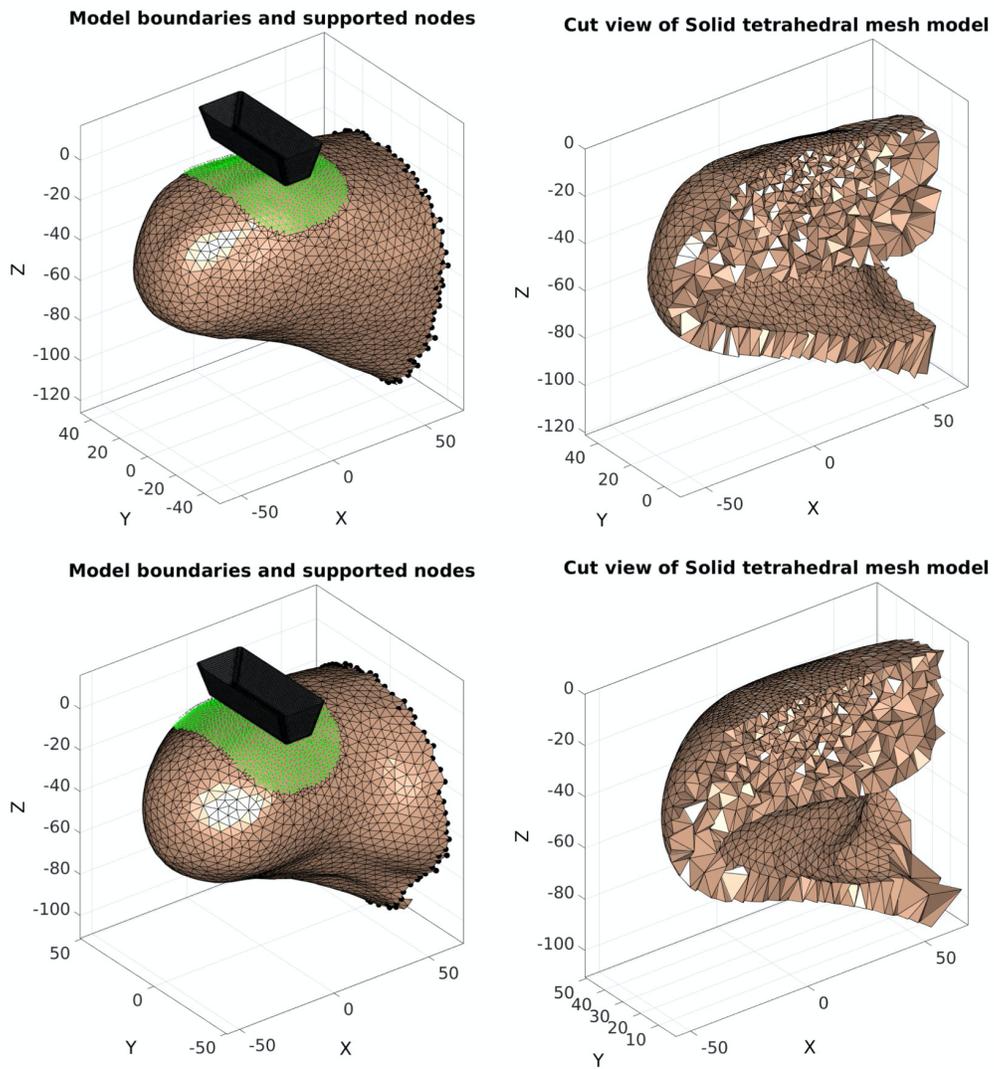

Figure 4: *FE meshes of the anatomically accurate local computational model created from surfaces of bones and skin segmented from the MRI data. This allows for simulation of the ultrasound indentation. (Top) Subject 2, leg 1; (Bottom) Subject 2, leg 2.*

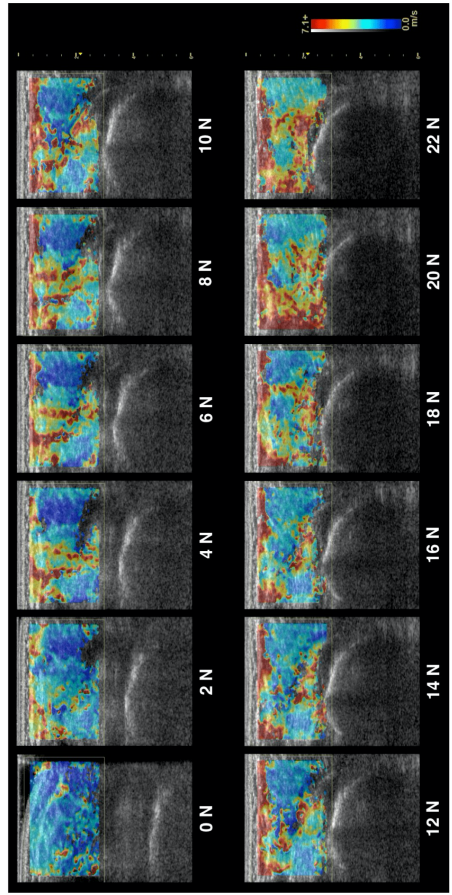

Figure 5: *Example sequence of ultrasound shear wave imagery collected at 2N incremental forces at a single indentation point on a residuum. At the 0 N unloaded state, the probe is just touching the limb and is not deforming the tissue, as demonstrated by the visible curvature of the skin. The shear wave velocity map increases in value as force increases. There is a noticeably higher increase in shear wave speeds in the region just above the bone, as compared to the surrounding tissue; this demonstrates that local stiffness measurements may be acquired during an indentation experiment using SWE.*

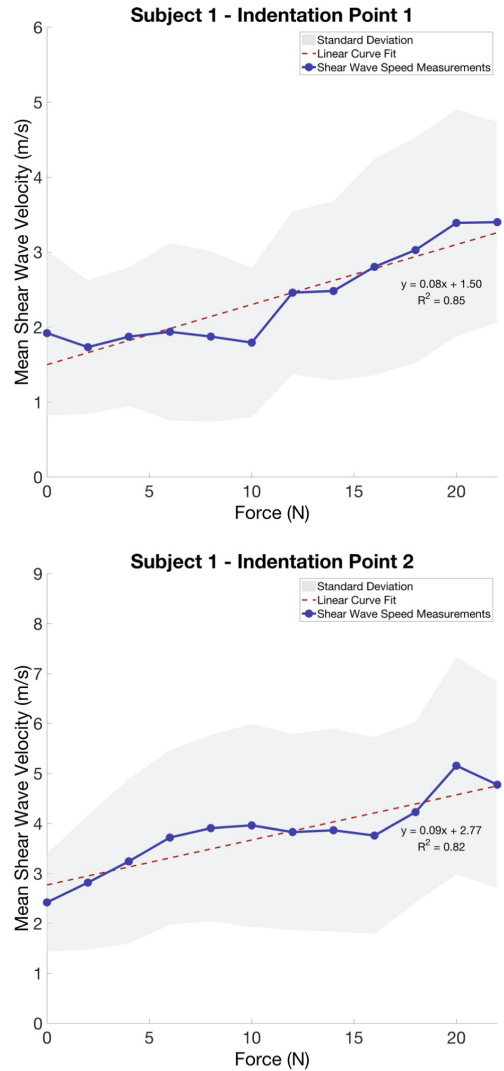

Figure 6: *Plots of mean shear wave speed at 2 N incremental forces at the two indentation sites for Subject 1 analyzed in this study. As noted by the linear curve fits, each indentation showed a linear increase in shear wave speed as a function of applied transducer force. Standard deviations also increased in most cases as a function of force.*

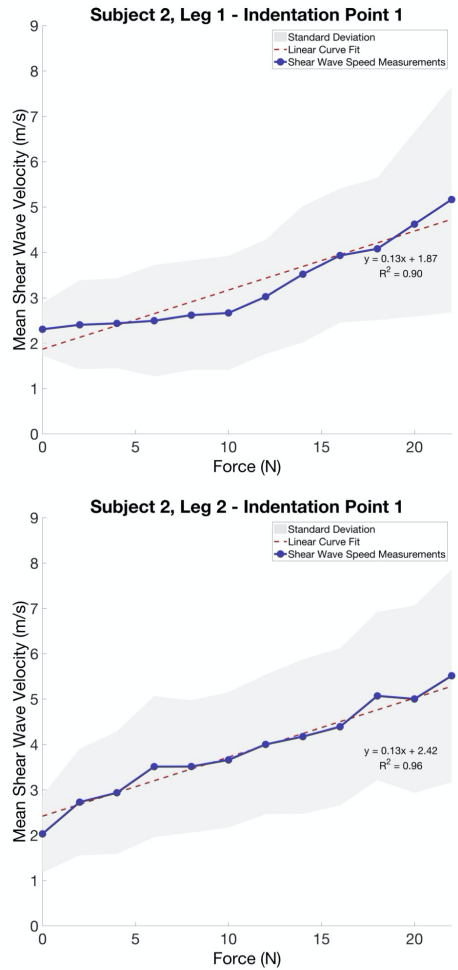

Figure 7: *Plots of mean shear wave speed at 2 N incremental forces at the two indentation sites for Subject 2 analyzed in this study. As noted by the linear curve fits, each indentation showed a linear increase in shear wave speed as a function of applied transducer force. Standard deviations also increased in most cases as a function of force.*

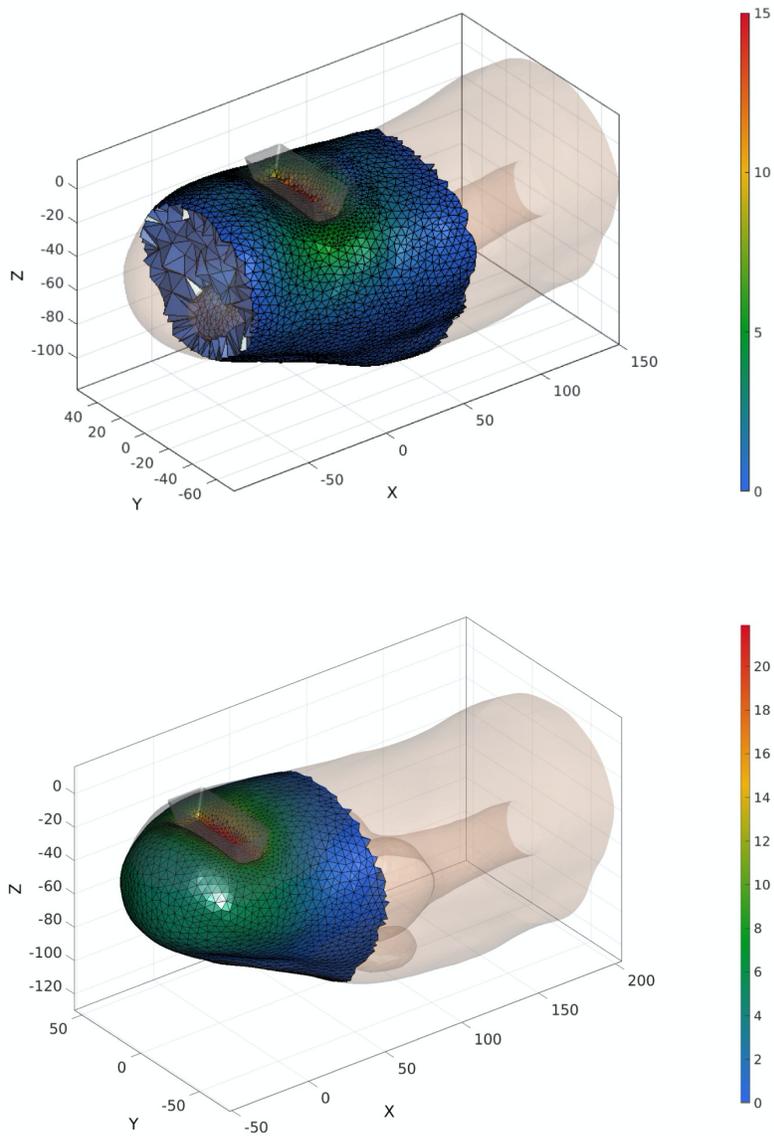

Figure 8: *Inverse FEA, showing the FE models of the indentation sites with tissue deformation during indentation. (Top) Subject 1, indentation point 1; (Bottom) Subject 1, indentation point 2.*

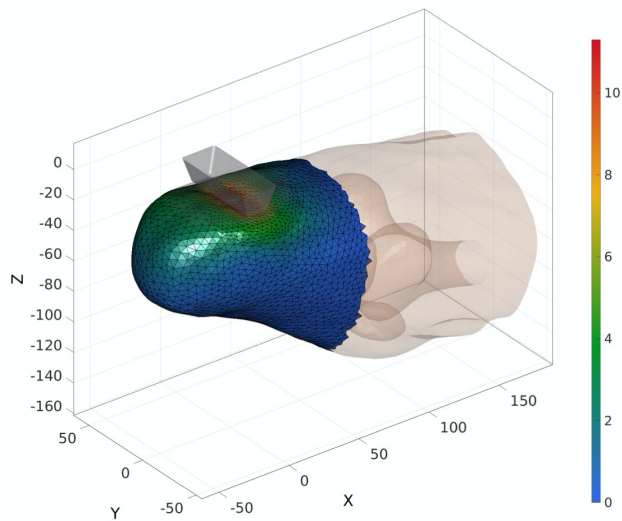

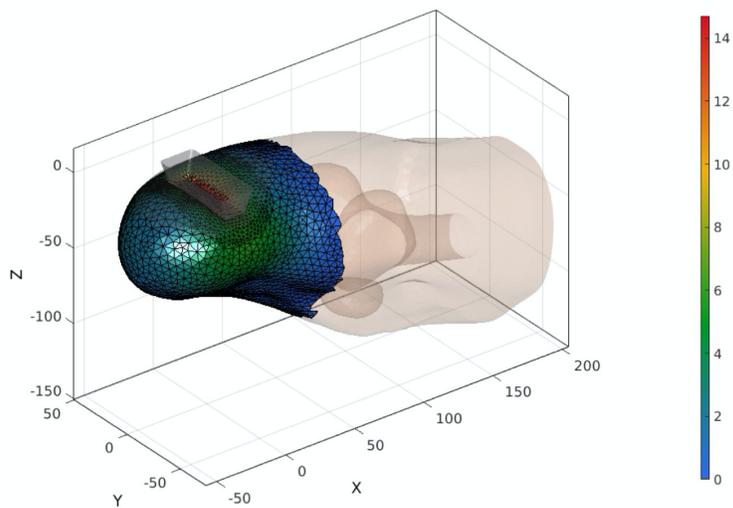

Figure 9: *Inverse FEA, showing the FE models of the indentation sites with tissue deformation during indentation. (Top) Subject 2, leg 1; (Bottom) Subject 2, leg 2.*

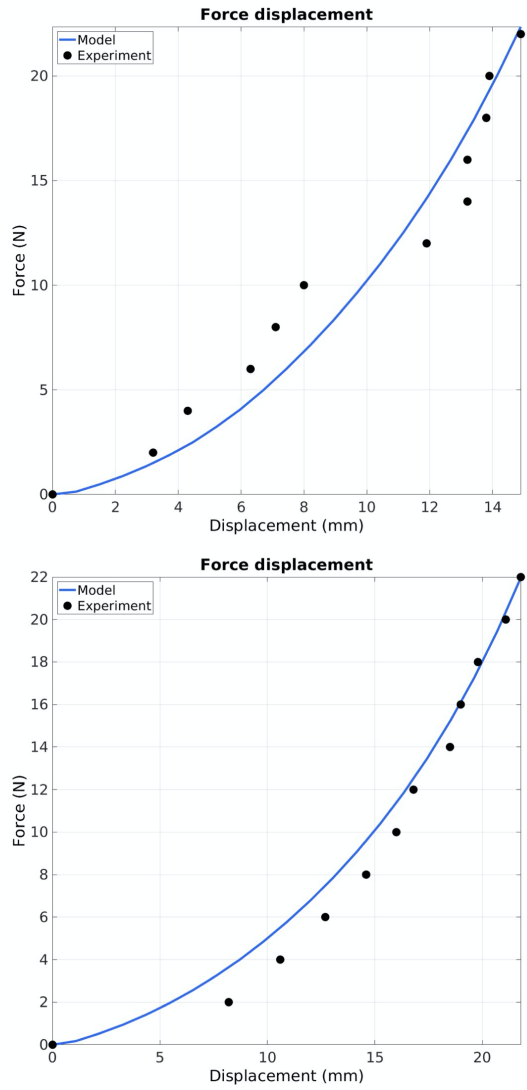

Figure 10: *FEA results showing force-displacement curves from the raw data and the resulting curve fit from inverse FEA. (Top) Subject 1, indentation point 1; (Bottom) Subject 1, indentation point 2.*

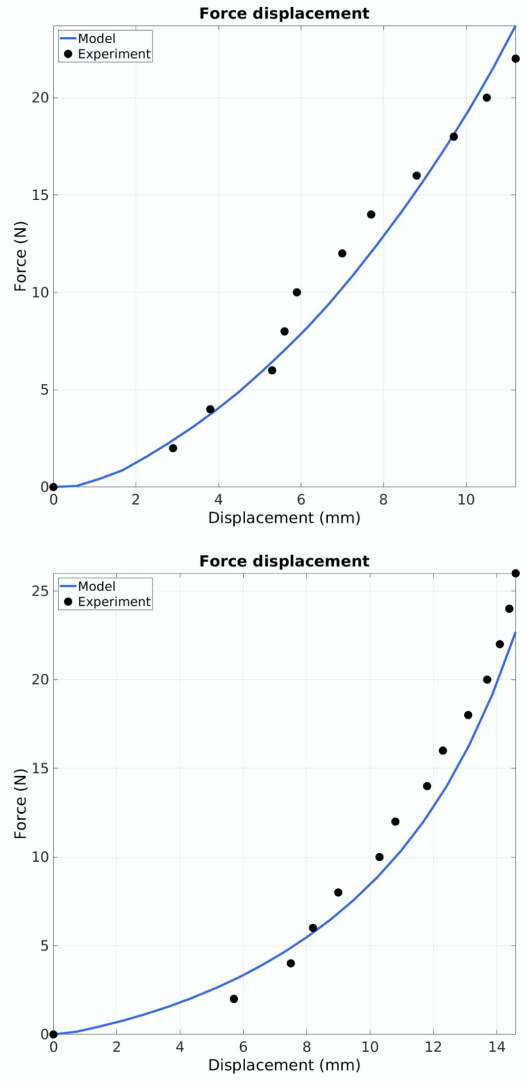

Figure 11: *FEA results showing force-displacement curves from the raw data and the resulting curve fit from inverse FEA. (Top) Subject 2, leg 1; (Bottom) Subject 2, leg 2.*

Table 1: Constitutive parameters determined from combined inverse FEA and shear wave elastography (SWE), along with associated correlation coefficient ($R^2$) for subject 1, indentation point 2.

| **Subject 1** | $c$ (kPa) | $m$ (.) | $\kappa'$ (kPa) | $R^2$ |
|---|---|---|---|---|
| **Mooney-Rivlin type** | 5.445 | 2.000 | 535.430 | 0.939 |
| **Ogden type** | 5.445 | 1.389 | 616.151 | 0.942 |

Table 2: Constitutive parameters determined from combined inverse FEA and shear wave elastography (SWE), along with associated correlation coefficient ($R^2$) for subject 2 (left).

| **Subject 2 (left)** | $c$ (kPa) | $m$ (.) | $\kappa'$ (kPa) | $R^2$ |
|---|---|---|---|---|
| **Mooney-Rivlin type** | 10.488 | 2.000 | 6723.570 | 0.973 |
| **Ogden type** | 10.488 | 1.044 | 6903.800 | 0.974 |

Table 3: Constitutive parameters determined from combined inverse FEA and shear wave elastography (SWE), along with associated correlation coefficient ($R^2$) for subject 2 (right).

| **Subject 2 (right)** | $c$ (kPa) | $m$ (.) | $\kappa'$ (kPa) | $R^2$ |
|---|---|---|---|---|
| **Mooney-Rivlin type** | 7.841 | 2.000 | 1370.950 | 0.945 |
| **Ogden type** | 7.841 | 3.245 | 1147.590 | 0.957 |

Frauziols, F., Chassagne, F., Badel, P., Navarro, L., Molimard, J., Curt, N., Avril, S., 2016. In vivo Identification of the Passive Mechanical Properties of Deep Soft Tissues in the Human Leg. Strain 52 (5), 400–411.

Gailey, R., Allen, K., Castles, J., Kucharik, J., Roeder, M., 2008. Review of secondary physical conditions associated with lower-limb amputation and long-term prosthesis use. Journal of rehabilitation research and development 45 (1), 15–29.

Guertler, C. A., Okamoto, R. J., Ireland, J. A., Pacia, C. P., Garbow, J. R., Chen, H., Bayly, P. V., 2020. Estimation of anisotropic material properties of soft tissue by mri of ultrasound-induced shear waves. Journal of Biomechanical Engineering 142 (3), 031001.

Henrot, P., Stines, J., Walter, F., Martinet, N., Paysant, J., Blum, A., 2000. Imaging of the painful lower limb stump. Radiographics 20, S219–S235.

Herbert, N., Simpson, D., Spence, W. D., Ion, W., 2005. A preliminary investigation into the development of 3-D printing of prosthetic sockets. The Journal of Rehabilitation Research and Development 42 (2), 141.

Hou, Z., Guertler, C. A., Okamoto, R. J., Chen, H., Garbow, J. R., Kamilov, U. S., Bayly, P. V., 2022. Estimation of the mechanical properties of a transversely isotropic material from shear wave fields via artificial neural networks. Journal of the Mechanical Behavior of Biomedical Materials 126, 105046.

Hou, Z., Okamoto, R. J., Bayly, P. V., 2020. Shear wave propagation and estimation of material parameters in a nonlinear, fibrous material. Journal of biomechanical engineering 142 (5).35 lower-extremity amputations. The Journal of bone and joint surgery. American volume 92 (17), 2852–68.

Wang, A. B., Perreault, E. J., Royston, T. J., Lee, S. S., 2019. Changes in shear wave propagation within skeletal muscle during active and passive force generation. Journal of biomechanics 94, 115–122.

Zheng, Y.-P., Mak, A. F., 1996. An ultrasound indentation system for biomechanical properties assessment of soft tissues in-vivo. IEEE transactions on biomedical engineering 43 (9), 912–918.40